\documentclass[aps,pre,twocolumn,superscriptaddress, nofootinbib]{revtex4-1}

\usepackage{xcolor,hyperref}
\hypersetup{
   colorlinks,
   linkcolor={blue!50!black},
   citecolor={blue!50!black},
   urlcolor={blue!80!black}
}

\usepackage{amsmath}
\usepackage{amssymb}
\usepackage{color}
\usepackage{numprint}
\usepackage{physics}
\usepackage{bm}
\usepackage{graphicx}
\usepackage{changes}
\usepackage{ bbold }
\usepackage{booktabs}
\usepackage{mathtools}
\usepackage{placeins}
\usepackage[ragged]{sidecap}
\usepackage{cleveref}

\newcolumntype{C}{>{$}c<{$}}
\AtBeginDocument{
\heavyrulewidth=.08em
\lightrulewidth=.05em
\cmidrulewidth=.03em
\belowrulesep=.65ex
\belowbottomsep=0pt
\aboverulesep=.4ex
\abovetopsep=0pt
\cmidrulesep=\doublerulesep
\cmidrulekern=.5em
\defaultaddspace=.5em
}

\DeclareMathAlphabet{\mathitbf}{OML}{cmm}{b}{it}

\newcommand{\dbar}{{\,\mathchar'26\mkern-12mu d}}

\newcommand{\sFrac}[2]{{\textstyle\frac{#1}{#2}}}
\DeclareMathOperator{\doubleCdot}{\vcentcolon}
\DeclareMathOperator{\quadCdot}{\vcentcolon\hspace*{0em}\vcentcolon}
\DeclareMathOperator{\tripleCdot}{\vcentcolon\hspace*{-0.3em}\cdot}
\newcommand{\plotlabel}[1]{(#1)}
\newcommand{\geert}[1]{#1}

\newcommand{\aref}[1]{Appendix~\ref{#1}}

\newcommand{\mode}[1]{\vb*{#1}}
\newcommand{\unitmode}[1]{\vu*{#1}}
\newcommand{\dyn}{\vb*{\mathcal{M}}}

\newcommand{\identity}{\mode{\mathcal{I}}}
\newcommand{\barrier}{V_{\text{B}}}

\begin{document}

\title{Nonlinear quasilocalized excitations in glasses. I. True representatives of soft spots}

\author{Geert Kapteijns}
\affiliation{Institute for Theoretical Physics, University of Amsterdam, Science Park 904, Amsterdam, Netherlands}
\author{David Richard}
\affiliation{Institute for Theoretical Physics, University of Amsterdam, Science Park 904, Amsterdam, Netherlands}
\affiliation{Department of Physics, Syracuse University, Syracuse, NY 13244}
\author{Edan Lerner}
\affiliation{Institute for Theoretical Physics, University of Amsterdam, Science Park 904, Amsterdam, Netherlands}

\begin{abstract}
	Structural glasses formed by quenching a melt  possess a population of soft quasilocalized excitations --- often called `soft spots' --- that are believed to play a key role in various thermodynamic, transport and mechanical phenomena. Under a narrow set of circumstances, quasilocalized excitations assume the form of vibrational (normal) modes, that are readily obtained by a harmonic analysis of the multi-dimensional potential energy. In general, however, direct access to the population of quasilocalized modes via harmonic analysis is hindered by hybridizations with other low-energy excitations, e.g.~phonons. In this series of papers we re-introduce and investigate the statistical-mechanical properties of a class of low-energy quasilocalized modes --- coined here \emph{nonlinear quasilocalized excitations} (NQEs) --- that are defined via an anharmonic (nonlinear) analysis of the potential energy landscape of a glass, and do not hybridize with other low-energy excitations. In this first paper, we review the theoretical framework that embeds a micromechanical definition of NQEs. We demonstrate how harmonic quasilocalized modes hybridize with other soft excitations, whereas NQEs properly represent soft spots without hybridization. We show that NQEs' energies converge to the energies of the softest, non-hybridized harmonic quasilocalized modes, cementing their status as true representatives of soft spots in structural glasses. Finally, we perform a statistical analysis of the mechanical properties of NQEs, which results in a prediction for the distribution of potential energy barriers that surround typical inherent states of structural glasses, as well as a prediction for the distribution of local strain thresholds to plastic instability.
\end{abstract}


\maketitle

\section{Introduction}\label{introduction}
A major goal in the past and current investigations of structural glasses is revealing and understanding the statistical-mechanical properties of low-energy excitations \cite{soft_potential_model_1991,Schober_prb_1992,Gurevich2003,Schober_Laird_numerics_PRL,schober1993_numerics,Schober_Oligschleger_numerics_PRB,ohern2003,Schober_Ruocco_2004,barrat_3d,vincenzo_epl_2010,mw_EM_epl,eric_boson_peak_emt,modes_prl,SciPost2016,protocol_prerc,modes_prl_2018,LB_modes_2019,ikeda_pnas,pinching_2019}, and the role these excitations play in various thermodynamic \cite{Anderson,Phillips,Zeller_and_Pohl_prb_1971}, mechanical \cite{lemaitre2004,micromechanics2016}, transport \cite{Zeller_and_Pohl_prb_1971,Ikeda_scattering_2018,scattering_jcp} and yielding phenomena \cite{tanguy2010,manning2011,rottler_normal_modes,plastic_modes_prerc,zohar_prerc}, as well as their connection to dynamics in the viscous supercooled liquid state \cite{Schober_correlate_modes_dynamics,widmer2008irreversible,harrowell_2009}.

It is now well-accepted that, at the lowest energies, glasses feature two types of excitations. Firstly, there are elastic waves (phonons), that must emerge due to the translational invariance of the potential energy \cite{sethna2006statistical}, and are predicted within (linear) continuum elasticity theory. While phonons in glasses are imperfect due to the structural and mechanical disorder \cite{phonon_widths}, the density of phonons of frequency $\omega$ nevertheless follows the Debye prediction $D(\omega) \sim \omega^{\dbar-1}$ in $\dbar$ spatial dimensions \cite{kittel2005introduction}. 

Secondly, glasses feature a population of soft, non-phononic quasilocalized modes (QLMs) that arises due to glasses' microscopic disorder and mechanical frustration, as established decades ago using computer simulations \cite{Schober_Laird_numerics_PRL,schober1993_numerics,Schober_Oligschleger_numerics_PRB}, and also argued for on theoretical grounds \cite{soft_potential_model_1991,Schober_prb_1992,Gurevich2003}. In computer glasses, these excitations are cleanly revealed as harmonic vibrational (normal) modes (see definition below) either below \cite{modes_prl} or between \cite{SciPost2016,ikeda_pnas} phonon bands, namely in the absence of strong hybridizations. In these circumstances, the non-phononic modes follow a universal gapless density of states $\mathcal{D}(\omega) \sim  \omega^4$ in the limit $\omega \to 0$ \cite{modes_prl,ikeda_pnas}, and are \emph{quasilocalized} --- they consist of a localized, disordered core, dressed by an Eshelby-like algebraic decay away from the core \cite{modes_prl}. The universal law $\mathcal{D}(\omega) \sim \omega^4$ was shown to be independent of spatial dimension \cite{modes_prl_2018}, microscopic details \cite{modes_prl}, and glass preparation protocol (e.g.~cooling rate) \cite{protocol_prerc, lerner2019finite}. The universal distribution even persists in the most deeply supercooled computer glasses \cite{pinching_2019,LB_modes_2019}.

It is precisely this population of soft QLMs, and their interaction with phonons, that is thought to influence various unexplained universal phenomena that are specific to structural glasses \cite{Anderson,Phillips,Zeller_and_Pohl_prb_1971,scattering_jcp}. Furthermore, a subset of QLMs were shown to represent the carriers of plasticity in externally-loaded glasses \cite{plastic_modes_prerc,lte_pnas,zohar_prerc}, and might be key in determining relaxation patterns in supercooled liquids \cite{Schober_correlate_modes_dynamics,widmer2008irreversible,harrowell_2009}. Therefore, knowledge of their full distribution is crucial for advancing our fundamental understanding of many glassy phenomena.

However, as noted above, in a normal mode analysis, localized soft spots assume the form of harmonic vibrational modes only when hybridizations with other low-frequency excitations --- in particular with phonons \cite{Schober_Ruocco_2004} --- do not occur. It has been shown in \cite{phonon_widths} that, in the thermodynamic limit, phonons dominate glasses' low-frequency spectra, and hybridizations of QLMs and phonons is inevitable. This makes it impossible to cleanly observe QLMs assuming the form of harmonic modes in that limit, which is a severe limitation of the harmonic analysis as a useful investigative tool of soft QLMs. We emphasize that the existence of QLMs --- as anomously soft spots in the glass --- and their effect on various physical phenomena, is \emph{not} dependent on whether they can be realized as harmonic modes or not (as shown e.g.~in \cite{lte_pnas}).

\begin{figure}[htpb]
	\centering
	\includegraphics[]{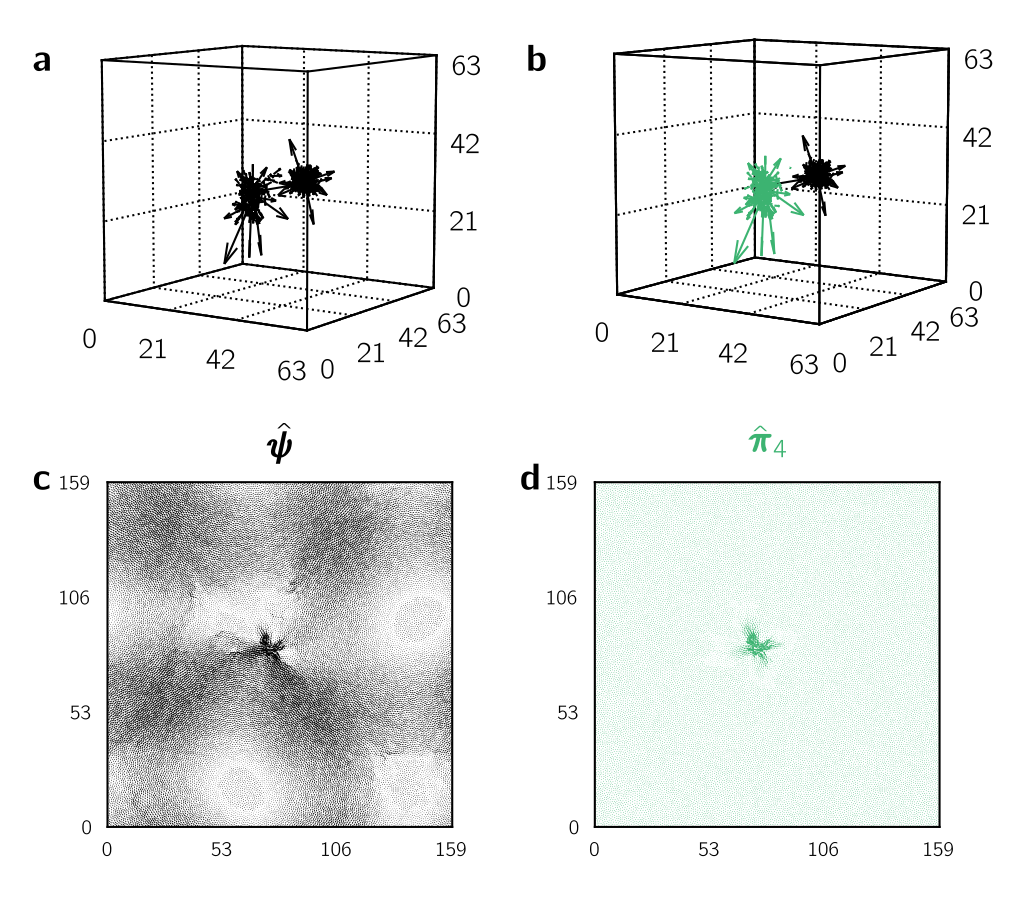}
	\caption{NQEs are indifferent to hybridization with excitations of nearby frequencies. Panel \plotlabel{a} shows a normal mode that consists of two hybridized quasilocalized excitations of similar frequency. In panel \plotlabel{b}, we show that these excitations are cleanly separated as two NQEs. Panel \plotlabel{c} shows a normal mode that is a mixture of a quasilocalized excitation and a nearby phononic excitation; panel \plotlabel{d} shows the dehybridized NQE. In panels \plotlabel{a} and \plotlabel{b}, only the largest 0.1\% of components of each mode is shown. For visual purposes, the modes are scaled so that the largest component of the NQE is the same size as the corresponding component of the harmonic mode. }
	\label{fig:hybridization}
\end{figure}

Recently, a theoretical framework was introduced \cite{SciPost2016} (which will be summarized in \autoref{sec:theoretical} below) that formulates quasilocalized excitations in a \emph{nonlinear} way --- outside of the normal (linear) mode analysis --- such that they are \emph{indifferent} to hybridizations with phonons or with other quasilocalized excitations of nearby frequencies. The robustness of these \emph{nonlinear quasilocalized excitations} (NQEs) against hybridizations is demonstrated in \autoref{fig:hybridization}; in the left panels we show low-frequency \emph{harmonic} modes measured in generic computer glasses, which consist of 
\emph{two} hybridized localized soft spots (panel \plotlabel{a}), and a localized soft spot hybridized with a phonon (panel \plotlabel{c}). In panel \plotlabel{b} we show that \emph{each} of the hybridized soft spots from panel \plotlabel{a} is represented by a \emph{different} NQE, while panel \plotlabel{d} shows that the soft spot that is hybridized with a phonon in panel \plotlabel{c} is represented by a single NQE, with no wave-like background (see \Cref{app:nqe} for details about the calculation of normal modes and NQEs). 

In this series of papers we present a comprehensive study of the statistical, mechanical, and statistical-mechanical properties of NQEs, together with a new methodological framework that enables the calculation of NQEs in generic computer glasses. We also extensively discuss the implications of our findings towards understanding some of the aforementioned fundamental and unresolved problems in glass physics.

In this first paper, in \autoref{sec:theoretical} we re-introduce --- in simple, physically transparent terms --- the theoretical framework from which a micromechanical definition of NQEs emerges. We then demonstrate in \autoref{sec:true} that NQEs are true representatives of soft spots in structural glasses: we show that NQEs disentangle localized soft spots from hybridized harmonic modes, and discuss in detail the energetic properties of NQEs compared to their harmonic counterparts. In \autoref{sec:cubic}, we put forward several key observations regarding the mechanical properties of NQEs, which enable important predictions about two fundamental properties of structural glasses: $(i)$ glasses' destabilization under externally-imposed deformations, and $(ii)$ the form of the distribution of potential-energy barriers that surround typical inherent states (local minima of the potential $U$) of a generic glass. Finally, in \autoref{sec:discussion} we discuss several avenues for future investigations, and look ahead to the other papers in this series. Our notation conventions are explained in \Cref{app:tensor}, and we describe in detail how NQEs are calculated in \Cref{app:nqe}.



\section{Computer glass models}\label{sec:models}
In this work, we employ two different glass-forming models. For most of the results reported below (and when not reported otherwise), data is shown for the inverse-power-law model, which we refer to as the IPL model from now on. It is a 50:50 binary mixture of `large' and `small' particles interacting with a pairwise potential $U(r) \sim r^{-10}$. The full details of this model, including microscopic units, elastic properties, cutoff radius and additional smoothing terms, are provided in \cite{boring_paper}. In 3D, we create glassy samples by performing an instantaneous quench from equilibrium liquid configurations at $T = 2.0$, which is 2.5 times larger than the temperature at which the athermal shear modulus of the underlying inherent states starts to saturate. We also study 2D and 4D glasses, prepared with the protocol described in \cite{modes_prl_2018}. We report the number of generated samples for each system size and spatial dimension in \autoref{tab:ipl}.

To study the effect of preparation protocol, we additionally employ a highly polydisperse inverse-power-law model \cite{LB_swap_prx}, that is optimized so that it can be extremely efficiently equilibrated with the Swap Monte Carlo algorithm \cite{tsai_swap, gazzillo_swap, grigera_swap}, while remaining robust against crystallisation. We use a slightly modified version of the model presented in \cite{LB_swap_prx}, the details of which are given in \cite{boring_paper}, which we refer to as the POLY model. We prepare glassy samples by performing an instantaneous quench from equilibrium liquid configurations at three different temperatures: $T = 0.35$ (deeply supercooled), $T = 0.5$ (moderately supercooled) and $T = 1.3$ (high temperature), which corresponds to approximately 40\%, 56\%, and 145\% of the temperature at which the athermal shear modulus of the underlying inherent states saturates.\footnote{In the units of the model presented in \cite{LB_swap_prx}, this corresponds approximately to $T = 0.06$, $T = 0.083$, and $T = 0.22$.} For each temperature, we create 10K samples of $N = 2000$, and 1K samples of $N = 8000$.

All quantities in this paper are reported in dimensionless microscopic units: lengths are rescaled by $a_0 \equiv (V/N)^{1/\dbar}$, and times (frequencies) are rescaled by $\tau_0 \equiv a_0 / c_{\text{T}}$ ($\omega_0 \equiv c_{\text{T}} / a_0$), where $c_{\text{T}} = \sqrt{G(T_{\text{p}} = \infty) / \rho}$ denotes the transverse wave speed, $G(T_{\text{p}} = \infty)$ the saturated (high-parent-temperature) athermal shear modulus \cite{boring_paper}, and $\rho$ the mass density. The values of $\rho$ and $G(T_{\text{p}} = \infty)$ for each model are reported in \autoref{tab:units}.

\begin{table}[]
\centering
\begin{tabular}{@{}lllllllll@{}}
\toprule
 & \multicolumn{4}{l}{2D IPL} & \multicolumn{3}{l}{3D IPL} & 4D IPL \\ \midrule
$N$ & 400 & 1600 & 6400 & 25600 & 2048 & 8192 & 32768 & 10000 \\
Ensemble size & 10K & 10K & 10K & 10K & 320K & 80K & 20K & 2K \\ \bottomrule
\end{tabular}
\caption{Ensemble size for each system size $N$ for the 2D IPL, 3D IPL, and 4D IPL models.  }
\label{tab:ipl}
\end{table}

\begin{table}[]
\centering
\begin{tabular}{@{}lllll@{}}
\toprule
 & 2D IPL & 3D IPL & 4D IPL & 3D POLY \\ \midrule
$G(T_{\text{p}} = \infty)$ & 15.8 & 12.4 & 10.9 & 9.22 \\
$\rho$ & 0.86 & 0.82 & 0.80 & 0.65 \\ \bottomrule
\end{tabular}
\caption{High-parent-temperature athermal shear modulus and mass density for the models employed in this paper.}
\label{tab:units}
\end{table}

\section{Theoretical framework}\label{sec:theoretical}
We consider a disordered system of $N$ particles in $\dbar$ spatial dimensions, enclosed in a fixed volume $V \equiv  L^\dbar$, and interacting via a potential $U(\mode{x})$, where $\mode{x}$ denotes the $N \dbar$-dimensional vector of all particles' coordinates. We assume all particle masses to be unity. For all analyses, we study the system in the zero-temperature limit, meaning that it resides in a local minimum $\mode{x}_0$ of the potential energy (also called an inherent state), so that all particles are in mechanical equilibrium --- the net force on each particle is zero.

In a conventional normal mode analysis, one expands the potential energy to second order in the displacement $\mode{z} \equiv \mode{x}  -  \mode{x}_0$ from this reference state as
\begin{equation}
	U(\mode{x}) - U(\mode{x}_0) \simeq \sFrac{1}{2} \dyn : \mode{z}\mode{z},
\end{equation}
where 
\begin{equation}
\dyn \equiv \eval{\pdv{U}{\mode{x}}{\mode{x}}}_{\mode{x}_0}
\end{equation}
is the Hessian matrix of the potential $U$. Conventionally, the $N\dbar$ eigenvectors $\unitmode{\psi}_{\ell}$ of $\dyn$ are referred to as \emph{normal modes} (also as \emph{vibrational modes} or \emph{harmonic modes}), with real eigenvalues $\omega^2_{\ell} \ge 0$ (because $\mode{x}_0$ represents a local minimum of $U$). The eigenvectors $\unitmode{\psi}_{\ell}$ represent orthogonal displacement directions along which the system can perform harmonic oscillatory motion with frequency $\omega_{\ell}$.

As mentioned above, structural glasses are known to feature soft quasilocalized excitations, which are displacement fields that consist of a localized, disordered core dressed by an algebraically-decaying field away from the core \cite{SciPost2016}. These excitations may be approximately realized as normal modes if there exist no other modes with similar frequency in the system. However, as soon as this is no longer the case, quasilocalized excitations are manifested as \emph{hybridized} normal modes, that mix a quasilocalized excitation with phonons or with other quasilocalized excitations, as demonstrated in \autoref{fig:hybridization}, and discussed at length in \cite{SciPost2016} and further below. Hybridization in the normal mode analysis can be understood as a direct consequence of the orthogonalization constraint on eigenmodes $\unitmode{\psi}_\ell$: quasilocalized excitations are \emph{not} perfectly orthogonal to phonons or other quasilocalized excitations, meaning that both phonons and quasilocalized excitations cannot be exact eigenmodes of the Hessian matrix, but must always be, to some extent, hybridized with each other. 

\subsection{Nonlinear quasilocalized excitations (NQE)}

We now present an alternative way to define quasilocalized excitations \emph{outside} of the conventional harmonic analysis of the potential energy. Consider a glass at zero temperature and at mechanical equilibrium, whose particles are subjected to an imposed displacement about the mechanical equilibrium state, of the form $\delta\mode{x} =  s\unitmode{z}$, i.e.~particles are displaced a distance $s$ in a general direction $\unitmode{z}$ on the multidimensional potential energy landscape. Since the initial state considered was a state of mechanical equilibrium, the system will respond to an imposed displacement with a restoring force, denoted here by $\mode{f}_{\unitmode{z}}(s)$. A third-order Taylor expansion of this force is written as
\begin{equation}\label{foo07}
	\mode{f}_{\unitmode{z}}(s) \simeq \mode{f}_1(\unitmode{z}) s + \mode{f}_2(\unitmode{z}) s^2 + \mode{f}_3(\unitmode{z}) s^3,
\end{equation}
where the coefficients of the expansion are given by
\begin{align}
	\mode{f}_1(\unitmode{z}) &= -\frac{\partial^2U}{\partial\mode{x}\partial\mode{x}}\cdot\unitmode{z} = - \dyn \cdot \unitmode{z}, \\
	\mode{f}_2(\unitmode{z}) &= - \frac{1}{2}\frac{\partial^3U}{\partial\mode{x}\partial\mode{x}\partial\mode{x}}\doubleCdot\unitmode{z}\unitmode{z} \equiv -\frac{1}{2}\mode{U}^{(3)} \doubleCdot \unitmode{z}\unitmode{z},\\
	\mode{f}_3(\unitmode{z}) &= - \frac{1}{6}\frac{\partial^4U}{\partial\mode{x}\partial\mode{x}\partial\mode{x}\partial\mode{x}}\tripleCdot  \unitmode{z} \unitmode{z} \unitmode{z} \equiv - \frac{1}{6} \mode{U}^{(4)} \tripleCdot \unitmode{z} \unitmode{z} \unitmode{z}\,.
\end{align}
Once the expansion of the force response $\mode{f}_{\unitmode{z}}(s)$ is spelled out, a mechanical interpretation of the definition of normal modes emerges; normal modes are collective displacement directions $\unitmode{\psi}$ for which $\mode{f}_1(\unitmode{\psi}) \propto  -\unitmode{\psi}$, i.e.~they are directions $\unitmode{\psi}$ for which the linear force response $\mode{f}_1(\unitmode{\psi})$ is \emph{parallel} to $-\unitmode{\psi}$ itself, implying that
\begin{equation}
\dyn\cdot\unitmode{\psi} = c_2 \unitmode{\psi},
\end{equation}
with $c_2$ a scalar, for which by definition we must have $c_2 = \omega^2$. The subscript `2' refers to the fact the modes $\unitmode{\psi}$ arise from a harmonic, i.e.~second-order expansion of the potential energy, in analogy with what follows.

What do we find if we instead consider collective displacement directions $\unitmode{\pi}_4$ whose linear force response $\mode{f}_1(\unitmode{\pi}_4)$ is parallel to the \emph{third-order} force response $\mode{f}_3(\unitmode{\pi}_4)$, i.e.~$\mode{f}_1(\unitmode{\pi}_4) \propto \mode{f}_3(\unitmode{\pi}_4)$? This would imply
\begin{equation}\label{eq:quartic_def}
	\dyn \cdot \unitmode{\pi}_4 = c_4 \mode{U}^{(4)} \tripleCdot \unitmode{\pi}_4 \unitmode{\pi}_4 \unitmode{\pi}_4.
\end{equation}
We coin these displacement directions $\unitmode{\pi}_4$ \emph{quartic modes}, because they arise from a fourth-order expansion of the potential energy.\footnote{\label{foot:cubic}A similar equation for \emph{cubic modes} may be derived by requiring $\mode{f}_2(\unitmode{\pi}_3) \propto \mode{f}_1(\unitmode{\pi}_3)$. We will discuss cubic modes extensively in \autoref{sec:cubic} of this paper.}
By contracting both sides of \autoref{eq:quartic_def} with $\unitmode{\pi}_4$, the proportionality constant $c_4$ is found to be
\begin{equation}
	c_4 = \frac{\kappa(\unitmode{\pi}_4)}{\chi(\unitmode{\pi}_4)},
\end{equation}
where we have defined the quartic expansion coefficient
\begin{equation}\label{eq:chi_def}
\chi(\mode{z}) \equiv \frac{\mode{U}^{(4)}\quadCdot\mode{z}\mode{z}\mode{z}\mode{z}}{(\mode{z}\cdot\mode{z})^2},
\end{equation}
and the stiffness $\kappa$ associated with a mode $\mode{z}$ as
\begin{equation}
\kappa(\mode{z}) \equiv \frac{\dyn\doubleCdot\mode{z}\mode{z}}{\mode{z}\cdot\mode{z}}.
\end{equation}
We note that for harmonic modes $\unitmode{\psi}_{\ell}$, the stiffness $\kappa(\unitmode{\psi}_{\ell}) = \omega_{\ell}^2$.

We next assert that solutions $\unitmode{\pi}_4$ to \autoref{eq:quartic_def}, are, in fact, quasilocalized excitations. To see this, consider the energy defined as\footnote{In \cite{SciPost2016}, the square of the energy $\mathcal{E}_4$ was called the `cost function' $\mathcal{G}_4$.}
\begin{equation}\label{eq:quartic_energy}
	\mathcal{E}_4(\mode{z}) = \kappa(\mode{z})^2/\chi(\mode{z}), 
\end{equation}
which is a function of a displacement field $\mode{z}$; local minima of this energy are precisely quartic modes $\unitmode{\pi}_4$ by virtue of \autoref{eq:quartic_def}, namely\footnote{Because $\mathcal{E}_4$ does not depend on the vector norm of its argument --- in consistence with the fact that modes $\mode{z}$ represent \emph{directions} in the potential energy landscape --- we can properly take the partial derivative with respect to each component of $\mode{z}$.}
\begin{equation}\label{eq:gradient}
	\eval{ \pdv{\mathcal{E}_4}{\mode{z}} }_{\mode{z}=\unitmode{\pi}_4} = \mode{0}.
\end{equation}
\geert{The energy function $\mathcal{E}_4$ has no straightforward physical interpretation; nevertheless,} local minima of $\mathcal{E}_4$ are soft, quasilocalized excitations, since they tend to have $(i)$ a small stiffness $\kappa$ (because it appears in the enumerator), and $(ii)$ a large quartic expansion coefficient $\chi$ (because it appears in the denominator), which, as we show now, means the mode is localized.  

\begin{figure}[h]
	\centering
	\includegraphics[]{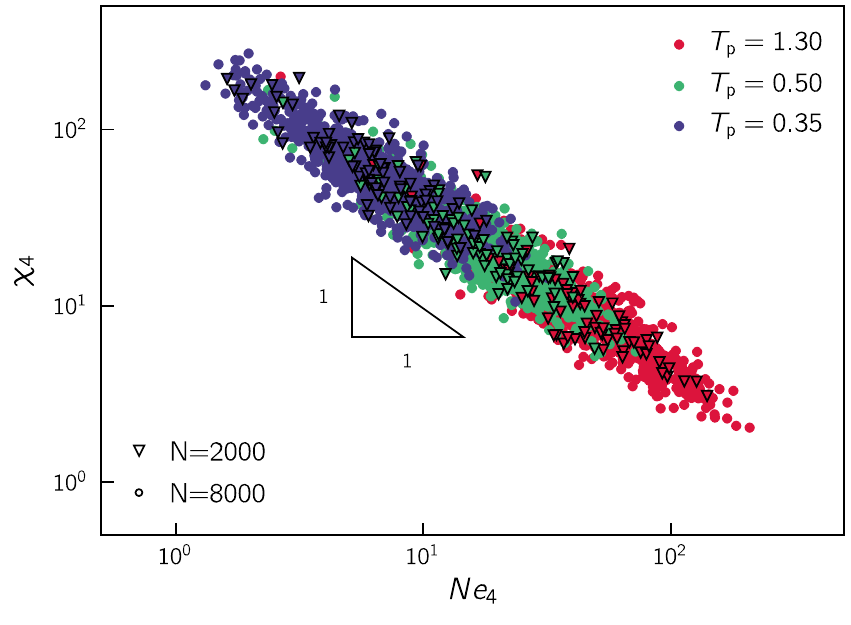}
	\caption{The quartic expansion coefficients $\chi_4$ are primarily sensitive to the localization $N e_4$ of the associated NQEs, independent of preparation protocol or system size of the glass.}
	\label{fig:chi_vs_e}
\end{figure}

To demonstrate how $\chi(\unitmode{\pi}_4)$ is correlated with the localization of $\unitmode{\pi}_4$, we scatter plot in \autoref{fig:chi_vs_e} $\chi_4 \equiv \chi(\unitmode{\pi}_4)$ vs.~$Ne_4$ (with $e_4 \equiv e(\unitmode{\pi}_4)$), where the participation ratio $e$ is defined as
\begin{equation}\label{eq:e_def}
e(\mode{z}) \equiv \frac{\big( \sum_i \mode{z}_i \cdot \mode{z}_i \big)^2}{N \sum_i \left( \mode{z}_i \cdot \mode{z}_i \right)^2}.
\end{equation}
Here, $\mode{z}_i$ denotes the $\dbar$-dimensional coordinate vector of the $i$th particle. The participation ratio of a field is a measure of its degree of localization: delocalized, extended modes feature $e \sim {\cal O}(1)$, whereas localized modes feature $e \sim {\cal O}(1/N)$. We find that $\chi_4 \sim (Ne_4)^{-1}$, as has also been shown in \cite{SciPost2016}. Remarkably, $\chi_4$ appears to be only sensitive to the degree of localization of $\unitmode{\pi}_4$, and is completely blind to the degree of annealing of the glass in which $\unitmode{\pi}_4$ is embedded. We present a rationalization for this general scaling relation in \Cref{app:chi_vs_Ne}.

The formulation of quartic modes, or \emph{quartic NQEs}, as local minima of an energy $\mathcal{E}_4$, allows for their numerical computation with standard minimization methods, as described in detail in \cite{SciPost2016} and in \Cref{app:nqe} of this work. 

\section{NQEs are true representatives of soft spots}\label{sec:true}

We will now establish that the NQEs defined in the previous section are true representatives of soft spots in model structural glasses. We show that NQEs do not hybridize with other low-energy excitations, and maintain their universal structure --- a disordered core dressed by an algebraically decaying far-field.\footnote{\geert{We note that it may be possible to design other cost functions that --- similar in spirit to $\mathcal{E}_4$ in \autoref{eq:quartic_energy} --- simultaneously minimize stiffness and maximize localization, and whose solutions also disentangle quasilocalized excitations from other low-energy excitations (see discussion in \Cref{app:chi_vs_Ne}). We will not focus on this possibility in the present work. 
}} The central piece of evidence supporting our claim is the fact that in the limit $\omega \to 0$ --- in the absence of hybridization --- harmonic QLMs converge to their corresponding NQEs both energetically and structurally. In this section, we quantitatively explain the observed convergence behaviour as a consequence of normal mode hybridization. A description of our protocol for finding the lowest harmonic QLM and NQE for each system is given in \Cref{app:nqe}. 

\begin{figure}[htpb]
	\centering
	\includegraphics[]{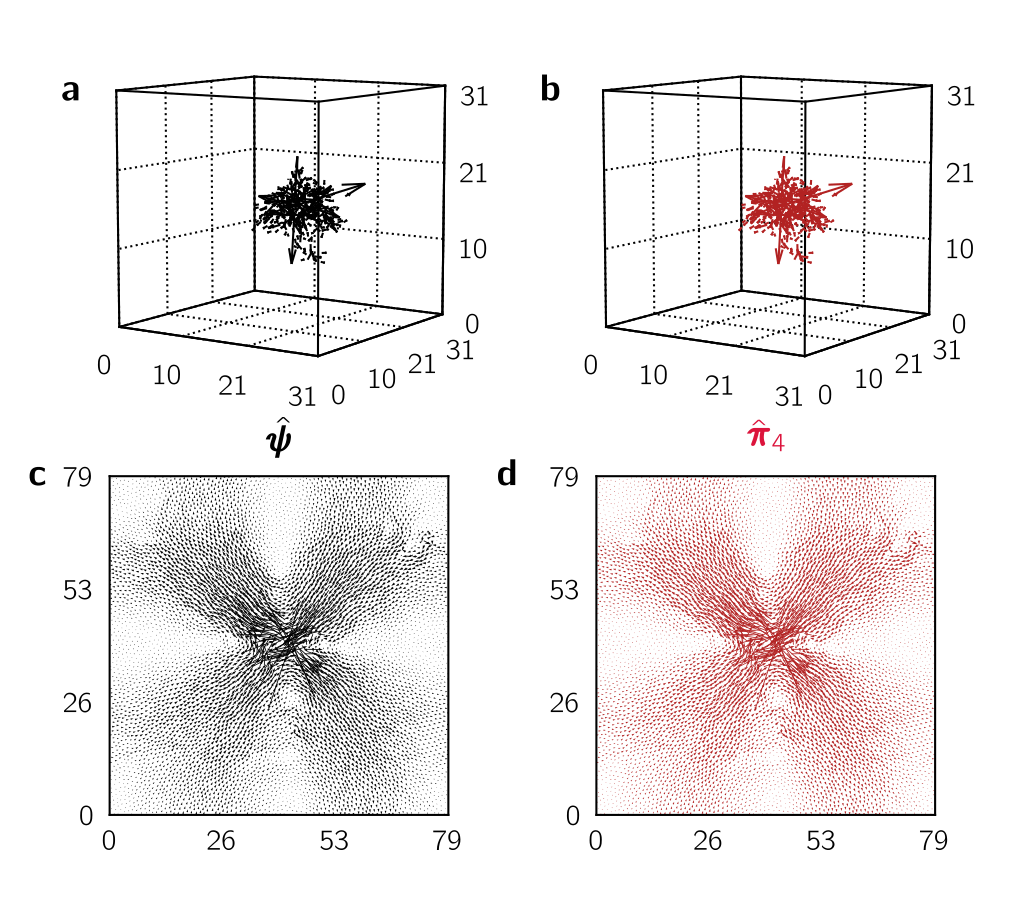}
	\caption{Harmonic QLM $\unitmode{\psi}$ and quartic NQE $\unitmode{\pi}_4$ converge energetically and structurally in the limit of low frequencies. Panels \plotlabel{a} and \plotlabel{c} show the softest harmonic QLM in the entire ensemble of solids of the IPL model in 3D and 2D. Panels \plotlabel{b} and \plotlabel{d} show their quartic nonlinear counterparts. The relative difference in energy between the harmonic and quartic excitations $(\omega_4^2 - \omega_2^2) / \omega_4^2$ is 0.15\% in 3D and 2.1\% in 2D; the structural difference $1 - |\unitmode{\psi} \cdot \unitmode{\pi}_4|$ is $3.6 \times 10^{-6}$ in 3D and $2.4 \times 10^{-4}$ in 2D.}
	\label{fig:clean_modes}
\end{figure}

\begin{figure}[htpb]
	\centering
	\includegraphics[]{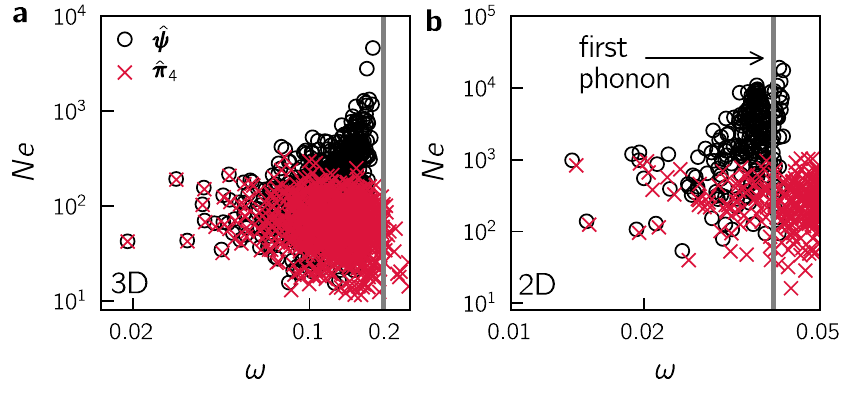}
	\caption{Every harmonic QLM $\unitmode{\psi}$ has a corresponding quartic NQE $\unitmode{\pi}_4$ of similar frequency and participation in 3D (panel \plotlabel{a}) and 2D (panel \plotlabel{b}). Because the harmonic QLM represents the direction of lowest stiffness in the entire system, the frequency of the NQE is always slightly higher, whereas the NQE --- which is free of hybridization --- has a slightly lower participation ratio. When the QLM's frequency approaches the frequency of the first phonon band, hybridization becomes stronger. This is reflected in a higher participation $Ne$ (and see also \autoref{fig:hybridization}(c)).}
	\label{fig:convergence_scatter}
\end{figure}

\begin{figure*}[htpb]
	\centering
	\includegraphics[]{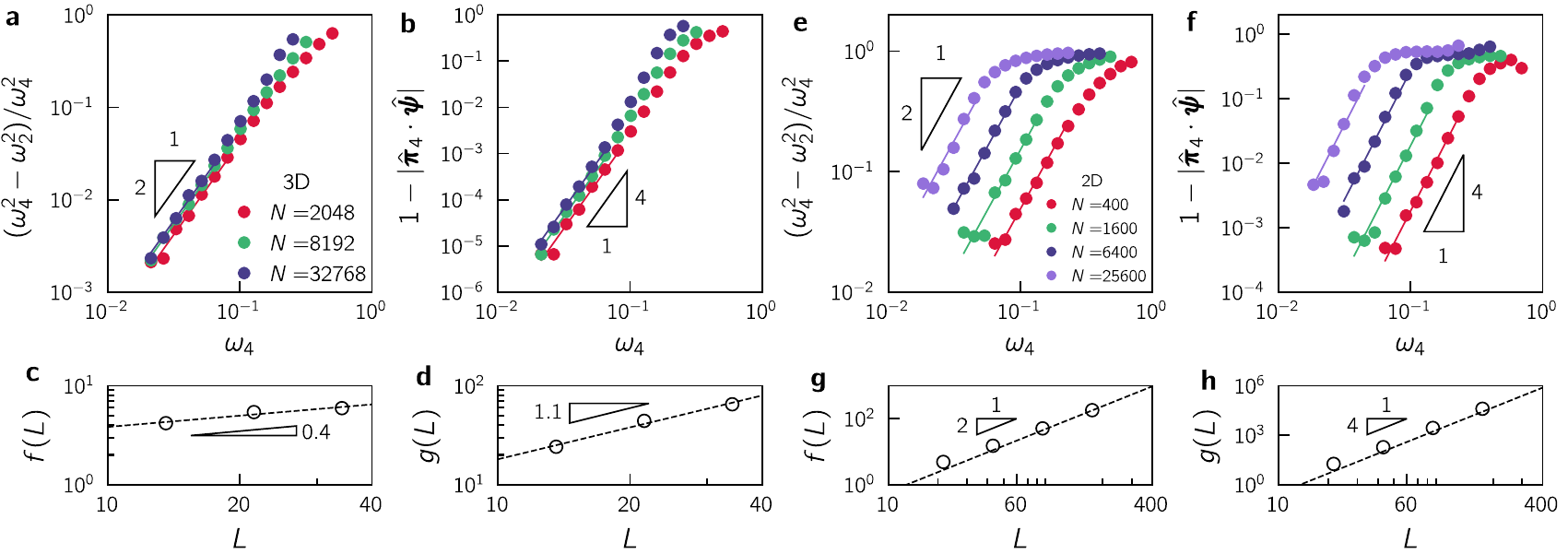}
	\caption{Harmonic QLM $\unitmode{\psi}$ and quartic NQE $\unitmode{\pi}_4$ converge energetically and structurally in the limit of low frequencies. Panels \plotlabel{a} and \plotlabel{e} show the energetic convergence (\autoref{eq:convergence_omega}) in 3D and 2D. Panels \plotlabel{b} and \plotlabel{f} show the structural convergence (\autoref{eq:convergence_overlap}) in 3D and 2D. Panels \plotlabel{c} and \plotlabel{d} show that the prefactors $f(L)$ and $g(L)$ of the energetic and structural convergence are weakly $L$-dependent in 3D, whereas panels \plotlabel{g} and \plotlabel{h} show they follow $f(L) \sim L^2$ and $g(L) \sim L^4$ in 2D. For panels (a, b, e, f), the data points represent the running median.}
	\label{fig:convergence}
\end{figure*}

We first visually demonstrate the convergence of harmonic QLMs and NQEs in \autoref{fig:clean_modes}, by comparing side-by-side the lowest-frequency quartic NQE of our entire ensemble of solids, together with its harmonic counterpart. We confirm that when hybridization is weak, the harmonic and quartic modes are nearly indistinguishable. In \autoref{fig:convergence_scatter}, we show that this holds for our entire ensemble of solids: every harmonic QLM has a nonlinear counterpart of very similar frequency and participation. In fact, because a harmonic QLM (below the first phonon band) by definition represents the direction of lowest energy of the system, the corresponding NQE always has a slightly \emph{higher} energy, and --- because it is not hybridized --- a \emph{lower} participation ratio. \autoref{fig:convergence}(a, e) shows that in the limit $\omega \to 0$, the relative difference in energy of these excitations scales as
\begin{equation}\label{eq:convergence_omega}
	(\omega_4^2 - \omega_2^2) / \omega_4^2 \sim f(L) \omega_4^2,
\end{equation}
both in 2D and 3D. Here, $\omega_2$ and $\omega_4$ denote the frequency of the harmonic and quartic excitations, respectively. We use the notation $\omega_4 \equiv \sqrt{\kappa(\unitmode{\pi}_4)}$ even though, strictly speaking, $\omega_4$ is not a vibrational frequency. $f(L)$ captures the $L$-dependence of the prefactor, which will be discussed further below; it increases weakly with $L$ in 3D, but strongly with $L$ in 2D. \autoref{fig:convergence}(b, f) shows that the structural convergence of the QLMs and NQEs --- measured by the difference of their overlaps from unity --- scales as
\begin{equation}\label{eq:convergence_overlap}
	1 - |\unitmode{\pi}_4 \cdot \unitmode{\psi}| \sim g(L) \omega_4^4,
\end{equation}
in 2D and 3D, where $g(L)$ again captures the $L$-dependence.

The $\omega_4$-dependence of the relations \autoref{eq:convergence_omega} and \autoref{eq:convergence_overlap} was shown in \cite{SciPost2016}, but no interpretation was given. In the following section, we quantitatively explain these scaling relations  --- including their $L$-dependence ---  as a consequence of harmonic QLMs' residual hybridization, even at very small frequencies, with other low-energy excitations in the system. We thus demonstrate that NQEs are \emph{true} quasilocalized soft spots, whereas harmonic QLMs in favorable circumstances approximate the NQEs, but are always to some degree `polluted' by other excitations due to the orthogonalization constraint of normal modes.

\subsection{Scaling argument for the convergence of QLMs and NQEs}

In this section, we present an argument quantitatively explaining the observed scaling relations \autoref{eq:convergence_omega} and \autoref{eq:convergence_overlap}.
Crucial to the discussion is the fact that any two non-degenerate normal modes must by definition be orthogonal, $\unitmode{\psi}_i \cdot \unitmode{\psi}_j = 0$. At low frequencies, excitations in the glass are either phonons or quasilocalized; an unhybridized quasilocalized excitation, which consists of a disordered core with an algebraically decaying far-field, is \emph{not} perfectly orthogonal to a phononic excitation or to other quasilocalized excitations. We conclude that in the normal mode analysis, both quasilocalized excitations and phonons cannot be perfectly represented, but must always be, to however small an extent, hybridized with each other.

For the analysis that follows, we assume the lowest harmonic mode of the system is a QLM, and its vibrational frequency is significantly lower than all other frequencies of the system, so that hybridization with other modes is weak. The central observable that we use to quantify the convergence of the harmonic QLM to its nonlinear counterpart is the difference vector $\mode{\Delta} \equiv \unitmode{\pi}_4 - \unitmode{\psi}$, which to first order in $|\mode{\Delta}|$ is given by
\begin{equation}\label{eq:delta}
	\mode{\Delta}  \simeq \omega_4^2 \sum_{\ell} \frac{\unitmode{\psi}_{\ell} \otimes \unitmode{\psi}_{\ell}}{\omega_{\ell}^2 - \omega_2^2} \left(\frac{\mode{U}^{(4)}\tripleCdot \unitmode{\pi}_4 \unitmode{\pi}_4 \unitmode{\pi}_4}{\chi_4} - \unitmode{\pi}_4 \right),
\end{equation}
where the sum runs over all normal modes $\unitmode{\psi}_{\ell}$ \emph{except} the lowest harmonic QLM $\unitmode{\psi}$. We derive this expression in \aref{app:scaling}. Since the magnitude of the field $\left(\mode{U}^{(4)}\tripleCdot \unitmode{\pi}_4 \unitmode{\pi}_4 \unitmode{\pi}_4 \right) / \chi_4$ is $O(1)$, independent of $\omega_4$ (see \autoref{fig:appendix2}\plotlabel{a} in \aref{app:data}), and the difference $\omega_{\ell}^2 - \omega_2^2$ is always large by assumption, it follows from this result that
\begin{equation}\label{eq:delta_scaling_with_omega}
|\mode{\Delta}| \sim \omega_4^2.
\end{equation}

\subsubsection{Structural convergence}
\begin{figure}[htpb]
	\centering
	\includegraphics[]{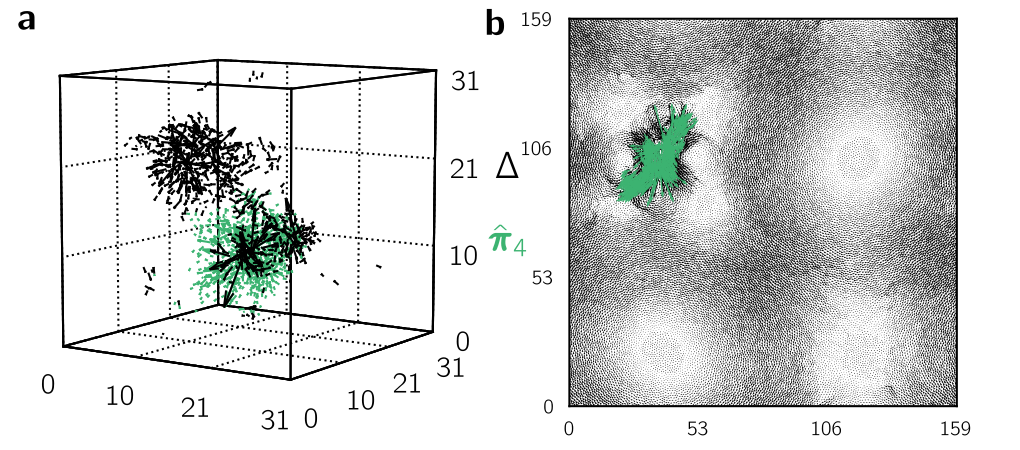}
	\caption{Representative examples of fields $\mode{\Delta} \equiv \unitmode{\pi}_4 - \unitmode{\psi}$ (see \autoref{eq:delta} and discussion in text) in (a) 3D and (b) 2D are shown in black. Shown in green are the corresponding NQEs $\unitmode{\pi}_4$. The magnitude of $\mode{\Delta}$ is much smaller than 1, so for visual purposes it has been scaled so that the size of the largest component of the core of $\mode{\Delta}$ matches the corresponding component of $\unitmode{\pi}_4$.
	For visualisation purposes, we show only the largest 2\% of components of $\mode{\Delta}$ and $\unitmode{\pi}_4$ in 3D, and the largest 1.4\% of $\unitmode{\pi}_4$ in 2D.}
	\label{fig:delta}
\end{figure}

We now focus on the structural convergence of the harmonic QLM $\unitmode{\psi}$ to the quartic NQE $\unitmode{\pi}_4$. Using \autoref{eq:delta_scaling_with_omega}, we write
\begin{equation}\label{eq:structural_conv}
	\frac{\mode{\Delta} \cdot \mode{\Delta}}{2} = 1 - \unitmode{\pi}_4 \cdot \unitmode{\psi} \sim \omega_4^4,
\end{equation}
which gives the correct $\omega_4$-dependence of the empirically observed \autoref{eq:convergence_overlap}. How can the $L$-dependence of the prefactor $g(L)$ be understood? In \autoref{fig:convergence}(h, d), we show that  
\begin{equation}\label{eq:g_2d}
	g(L) \sim L^4
\end{equation}
in 2D, whereas the dependence on $L$ is much weaker in 3D. To gain intuition for why the $L$-dependence is so strongly dimension-dependent, we visualize in \autoref{fig:delta} representative fields $\mode{\Delta}$ in 2D and 3D. $\mode{\Delta}$ always contains the disordered core of the soft quasilocalized excitation; however, it is precisely the far-field of $\mode{\Delta}$ that contains physically interesting information, as it reveals which excitations are hybridized with the harmonic QLM. In 2D, the concentration of QLMs below the lowest phonon band is lower than in 3D,\footnote{\label{foot:weibull}Since QLMs' frequencies are independently distributed according to $\mathcal{D}(\omega) \sim \omega^4$, a Weibullian scaling argument predicts that the frequency of the lowest QLM in a system of linear size $L$ follows $\langle \omega_{\text{min}} \rangle(L) \sim L^{-\dbar/5}$ \cite{modes_prl}. The lowest phonon frequency scales as $L^{-1}$, independent of dimension, explaining why --- all else being equal --- QLMs less frequently fall below the lowest phonon band in 2D than in 3D.} and therefore the hybridization with the lowest QLM is dominated by the phonons in the first phonon band, as the field $\mode{\Delta}$ in \autoref{fig:delta}(b) clearly shows. This implies that the sum in \autoref{eq:delta} is dominated by the first phonon band, whose frequency scales as $\omega_{\text{ph}} \sim L^{-1}$, from which it follows that $1 / (\omega_{\text{ph}}^2 - \omega_2^2) \sim L^2$. Hence,
\begin{equation}
	\mode{\Delta} \sim \omega_4^2 L^2  \left[ \unitmode{\psi}_{\text{ph}}  \cdot \left(\mode{U}^{(4)}\tripleCdot \unitmode{\pi}_4 \unitmode{\pi}_4 \unitmode{\pi}_4 / \chi_4 - \unitmode{\pi}_4 \right) \right] \unitmode{\psi}_{\text{ph}}.
\end{equation}
To obtain the full $L$-dependence of $\mode{\Delta}$, we next consider the overlap $\unitmode{\psi}_{\text{ph}}\cdot \left(\mode{U}^{(4)}\tripleCdot \unitmode{\pi}_4 \unitmode{\pi}_4 \unitmode{\pi}_4 / \chi_4 - \unitmode{\pi}_4 \right)$. The triple contraction $\left(\mode{U}^{(4)}\tripleCdot \unitmode{\pi}_4 \unitmode{\pi}_4 \unitmode{\pi}_4 \right) / \chi_4$ decays as $r^{-3\dbar}$ away from the core of the NQE \cite{SciPost2016}, and so its overlap with a phonon --- an extended field --- is small. The overlap with $\unitmode{\pi}_4$, which has a far-field decay of $r^{1 - \dbar}$, therefore dominates. Furthermore, the decay $r^{1 - \dbar}$ implies that --- because the surface of a sphere in $\dbar$ dimensions scales as $r^{\dbar - 1}$ --- the contraction of $\unitmode{\pi}_4$ with an extended field is system-size independent. These considerations together lead to the prediction that, if hybridization is due to the lowest phonons, $|\mode{\Delta}| \sim L^2 \omega^2_4$. Together with \autoref{eq:structural_conv} this explains \autoref{eq:g_2d}, and we conclude that in 2D, hybridization is indeed dominated by the lowest phonons. 

Turning to the 3D case, we observe that typical fields $\mode{\Delta}$ are \emph{not} wavelike, but instead mainly feature other QLMs in the far-field, as shown in \autoref{fig:delta}(a). This means that the sum in \autoref{eq:delta} is dominated by other QLMs \emph{below} the lowest phonon band. In the overlap between these QLMs and the field $\left(\mode{U}^{(4)}\tripleCdot \unitmode{\pi}_4 \unitmode{\pi}_4 \unitmode{\pi}_4 / \chi_4 - \unitmode{\pi}_4 \right)$, the term $\unitmode{\pi}_4$ again dominates. $\unitmode{\pi}_4$ couples only to QLMs in the vicinity, since these objects decay as $r^{1-\dbar}$ from their core; the frequency difference between the lowest QLM with the second-lowest, third-lowest, etc.~with non-negligible overlap is therefore only weakly dependent on $L$.\footnote{\label{foot:weibull2}The scaling argument discussed in \cref{foot:weibull} also applies to the second-lowest, third-lowest, etc.~frequency. Since QLMs have non-negligible overlap only with other QLMs in their vicinity, the frequency difference between the lowest QLM and the next-lowest in its vicinity is expected to depend very weakly on $L$. } Consequently, we expect $g(L)$ to be weakly $L$-dependent also, as we indeed observe.

\subsubsection{Energetic convergence}
Finally, we focus on the energetic convergence between the harmonic QLM $\unitmode{\psi}$ and the corresponding NQE $\unitmode{\pi}_4$, given by \autoref{eq:convergence_omega}. The energy of $\unitmode{\pi}_4 = \unitmode{\psi} + \mode{\Delta}$ is written as
\begin{equation}
	\omega_4^2 \equiv \dyn \doubleCdot \unitmode{\pi}_4\unitmode{\pi}_4 = \omega_2^2 + \dyn \doubleCdot \mode{\Delta}\mode{\Delta} + 2 \dyn \doubleCdot \unitmode{\psi}\mode{\Delta}.
\end{equation}
To rewrite the third term, we use the fact that $\dyn \cdot \unitmode{\psi} = \omega_2^2 \unitmode{\psi}$, and the result from the previous discussion $\unitmode{\psi} \cdot \mode{\Delta} = - |\mode{\Delta}|^2/2$ (cf.~\autoref{eq:structural_conv}). After rearranging, we obtain
\begin{equation}
	\omega_4^2 - \omega_2^2 = |\mode{\Delta}|^2 \dyn \doubleCdot \unitmode{\Delta}\unitmode{\Delta} -  |\mode{\Delta}|^3 \omega_2^2.
\end{equation}
$\dyn \doubleCdot \unitmode{\Delta}\unitmode{\Delta}$ represents the typical stiffness of the hybridized excitations, which by assumption is much larger than $\omega_2^2$. We can thus safely neglect the $O(|\mode{\Delta}|^3)$ term. As we discussed above, $|\mode{\Delta}|$ scales inversely proportional to the stiffness of the typical harmonic modes that hybridize with the lowest QLM, owing to the factor $(\omega_{\ell}^2 - \omega_2^2)^{-1} \sim (\dyn \doubleCdot \unitmode{\Delta}\unitmode{\Delta})^{-1}$ in the sum in \autoref{eq:delta}. Using \autoref{eq:delta_scaling_with_omega}, we obtain
\begin{equation}
	\frac{\omega_4^2 - \omega_2^2}{\omega_4^2} \sim \frac{\omega_4^2}{\dyn \doubleCdot \unitmode{\Delta}\unitmode{\Delta}}.
\end{equation}
In 2D, where hybridization is dominated by the lowest phonons, $\dyn \doubleCdot \unitmode{\Delta}\unitmode{\Delta} \sim \omega_{\text{ph}}^2 \sim L^{-2}$, so that
\begin{equation}
	f(L) \sim L^2,
\end{equation}
which is validated numerically in \autoref{fig:convergence}(g). In 3D, hybridization is dominated by nearby QLMs below the lowest phonon band, so that $f(L)$ scales weakly with $L$ (see \cref{foot:weibull2}). This is validated in \autoref{fig:convergence}(c).

In conclusion, we have quantitatively predicted the convergence of harmonic QLMs to their quartic counterparts as a consequence of the unavoidable hybridization of harmonic modes --- with phonons and/or other QLMs of similar frequencies --- strongly reinforcing our claim that quartic NQEs are true representatives of quasilocalized soft excitations in glasses.

\section{NQEs reveal universal properties of the potential energy landscape}\label{sec:cubic}
In the previous sections we focused on NQEs that are represented by solutions $\unitmode{\pi}_4$ to \autoref{eq:quartic_def}, referred to as quartic modes. In \autoref{sec:theoretical}, quartic modes where defined as displacement directions for which the linear force response is parallel to the third-order force response, $\mode{f}_1(\unitmode{\pi}_4) \propto \mode{f}_3(\unitmode{\pi}_4)$.
However, another important class of NQEs is that of displacement directions $\unitmode{\pi}_3$ for which the linear force response is parallel to the \emph{second-order} force response, $\mode{f}_1(\unitmode{\pi}_3) \propto \mode{f}_2(\unitmode{\pi}_3)$. Analogously to \autoref{eq:quartic_def}, this implies
\begin{equation}\label{cubic_modes_equation}
	\dyn \cdot \unitmode{\pi}_3 = c_3 \mode{U}^{(3)} \doubleCdot \unitmode{\pi}_3 \unitmode{\pi}_3.
\end{equation}
We refer to these modes as \emph{cubic modes}, because their definition requires a third-order expansion of the potential energy. By contracting both sides of \autoref{cubic_modes_equation} with $\unitmode{\pi}_3$, we find
\begin{equation}
	c_3 = \frac{\kappa(\unitmode{\pi}_3)}{\tau(\unitmode{\pi}_3)},
\end{equation}
where we have defined the cubic expansion coefficient
\begin{equation}
	\tau(\mode{z}) \equiv \frac{ \mode{U}^{(3)} \tripleCdot \mode{z} \mode{z} \mode{z} }{\left( \mode{z} \cdot \mode{z} \right)^{3/2}}.
\end{equation}

\geert{Cubic modes were first introduced in \cite{plastic_modes_prerc}, where they were shown to be very accurate representatives of the loci and geometry of imminent plastic instabilities that occur in sheared athermal glasses. In addition, the stiffnesses $\kappa_3 \equiv \kappa(\unitmode{\pi}_3)$ associated with cubic modes were shown to follow a simple dynamics with imposed shear strain $\gamma$, namely~\cite{micromechanics2016}
\begin{equation}\label{foo05}
	\dv{\kappa_3}{\gamma} \simeq - \frac{\tau_3 \nu_3}{\kappa_3},
\end{equation}
which becomes exact in the limit of small $\kappa_3$. Here $\nu_3 \equiv \frac{\partial^2 U}{\partial\gamma\partial\mode{x}} \cdot \unitmode{\pi}_3$ is the shear-strain-coupling parameter associated with the cubic mode $\unitmode{\pi}_3$. These properties of cubic modes suggest that they are promising candidates to represent the Shear Transformation Zones envisioned by Falk and Langer in their seminal work on plastic deformation in glasses \cite{falk_langer_stz}, as discussed in detail in \cite{micromechanics2016}.}

In this section we first discuss the energetic and statistical properties of cubic modes, and explain their differences and similarities compared to quartic modes. \geert{To this end, we study an ensemble of pairs of cubic and quartic modes that represent the \emph{same} soft spot in each of our computer glasses. The details of these calculations are presented in \Cref{app:nqe}.}
Finally, based on the unique mechanical properties of cubic modes, we make predictions about the universal statistical properties of generic computer glasses' potential energy landscape, and about their destabilization under shear.

\begin{figure*}[htpb]
	\centering
	\includegraphics[]{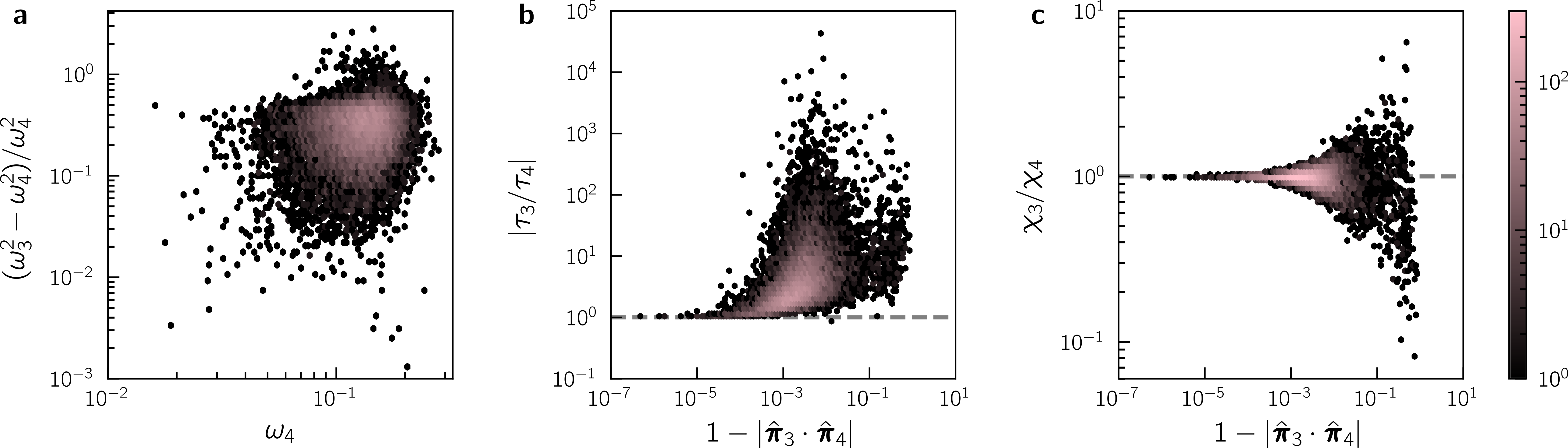}
	\caption{Difference between cubic and quartic NQEs. Panel (a) shows that the relative energy difference between cubic and quartic modes is approximately 30\%, independent of frequency, corresponding to a relative frequency difference of about 14\%. Panel (b) shows that the third-order coefficients $\tau_3, \tau_4$, which are a measure of the asymmetry of the energy landscape along the mode, are highly sensitive to the modes' structure, \geert{as quantified by the difference from unity of their overlap, $1 - |\unitmode{\pi}_3 \cdot \unitmode{\pi}_4|$.} Panel (c) shows that the fourth-order coefficients $\chi_3, \chi_4$ are insensitive to small changes in the modes' structure, consistent with the fact that $\chi$ scales inversely with the participation ratio (see \autoref{fig:chi_vs_e}), which is a global, geometric measure. We emphasize that in panel (b), the data span almost five orders of magnitude, whereas in panel (c), the data span less than one.  All data are reported as 2D histograms, with the bin counts given by the color of the bin.}
	\label{fig:why_cubic_modes}
\end{figure*}

\subsection{Cubic modes' energetics}\label{subsec:energetics}

In \autoref{fig:why_cubic_modes}\plotlabel{a} we plot the relative increase of cubic modes' stiffnesses $\omega_3^2$ over quartic modes stiffnesses $\omega_4^2$ (in our data, cubic modes' stiffnesses are larger than quartic modes' stiffnesses in over $99.5\%$ of cases). We find that cubic modes' stiffnesses are typically approximately 30\% higher than those of quartic modes that represent the same soft spots (corresponding to frequencies that are typically 14\% higher), independent of the frequency associated with those soft spots. In other words, we find that the \emph{relative} increase in frequency of cubic modes compared to quartic modes is constant, and independent of frequency. Since the distribution of quartic modes is expected to follow $\mathcal{D}(\omega_4) \sim \omega_4^4$ for small frequencies --- owing to their convergence to harmonic modes, which feature $\mathcal{D}(\omega_2) \sim \omega_2^4$ in the absence of strong hybridization \cite{modes_prl,ikeda_pnas,modes_prl_2018,pinching_2019,lerner2019finite} --- this means that the distribution of cubic modes is also expected to follow
\begin{equation}\label{foo00}
{\cal D}(\omega_3) \sim \omega_3^4\quad\mbox{as}\quad\omega_3\to 0,
\end{equation}
which will be key to additional results derived in what follows. 

In order to understand why the frequency of a cubic mode is generally higher than that of the quartic mode that represents the same soft spot, we plot in \autoref{fig:why_cubic_modes}(b) the absolute magnitude of the ratio of the third-order coefficients associated with cubic modes, to the third-order coefficients associated with quartic modes, $|\tau_3/\tau_4|$, against the difference from unity of their overlap $1 - |\unitmode{\pi}_3 \cdot \unitmode{\pi}_4|$. These data indicate that very minute differences in the modes' structure can lead to very large --- up to several orders of magnitude --- changes in their third-order coefficients. 

In contrast, the data presented in \autoref{fig:why_cubic_modes}(c) indicate that fourth-order coefficients $\chi_3,\chi_4$ are quite indifferent to small changes in modes' structure. This is expected, since we demonstrated in \autoref{fig:chi_vs_e} that fourth-order coefficients are primarily sensitive to the degree of localization of their associated modes, which is a purely geometric measure.
As discussed in \autoref{sec:theoretical}, quartic modes are defined by collective displacements that feature both small stiffnesses $\kappa_4 \equiv \omega_4^2$, \emph{and} large fourth-order coefficients $\chi_4$, as can be deduced from the `quartic energy' function (\autoref{eq:quartic_energy}), of which modes $\unitmode{\pi}_4$ are local minima. Having established that the coefficients $\chi_4$ associated with quartic modes are insensitive to details in the mode's structure, we conclude that maximizing $\chi_4$ does not impose large constraints on minimizing $\omega_4$.

For cubic modes, a similar `cubic energy' is given by\footnote{In \cite{SciPost2016}, $\mathcal{E}_3$ was called the `cost function' $\mathcal{G}_3$, and in \cite{plastic_modes_prerc, micromechanics2016} it was called the `barrier function' $b$.}
\begin{equation}\label{eq:cubic_energy}
	\mathcal{E}_3(\mode{z}) \equiv \frac{\kappa(\mode{z})^3}{\tau(\mode{z})^2}.
\end{equation}
Analogously to the scenario for quartic modes, cubic modes --- being local minima of $\mathcal{E}_3$ --- tend to have small stiffnesses $\kappa_3$, and large \emph{third-order} coefficients $\tau_3$. Since the coefficients $\tau_3$ are very sensitive to fine details of cubic modes' structure, as we have shown in \autoref{fig:why_cubic_modes}(b), we expect that maximizing $\tau_3$ imposes large constraints on cubic modes' frequencies $\omega_3$, which, in turn, explains why they are generally larger than quartic modes' frequencies.

\subsection{Cubic modes' stability}\label{subsec:stability}

Within the Soft Potential Model \cite{soft_potential_model_1991,Schober_prb_1992}, a stability argument is spelled out, according to which the the expansion coefficients $\kappa,\tau,\chi$ associated with soft QLMs satisfy the soft inequality
\begin{equation}\label{foo01}
\tau^2 \le 3\kappa\chi, 
\end{equation}
independent of glass stability.
This soft inequality implies that, within a quartic expansion of the potential energy along a soft QLM, if a second minimum exists --- meaning that the expansion represents a double-well potential --- the energy of the second potential well is necessarily \emph{larger} than the minimum in which the system resides. 

Based on the sensitivity of NQEs' third-order coefficients $\tau$ to the fine details of NQEs' structure, as demonstrated in the previous subsection (see Fig.~\ref{fig:why_cubic_modes}(b)), we argue that inequality~(\ref{foo01}) can only be meaningfully tested with cubic modes, since cubic modes' third-order coefficients $\tau_3$ are maximal, while still maintaining approximately the \emph{same frequencies as the softest NQEs} --- the quartic modes --- as shown in Fig.~\ref{fig:why_cubic_modes}(a). In other words, we assert that the third-order coefficients of quartic modes ($\tau_4$) or of harmonic QLMs ($\tau_2$) do \emph{not} represent the true asymmetry of soft spots, but, crucially, that $\tau_3$ does. 

\begin{figure}[htpb]
	\centering
	\includegraphics[]{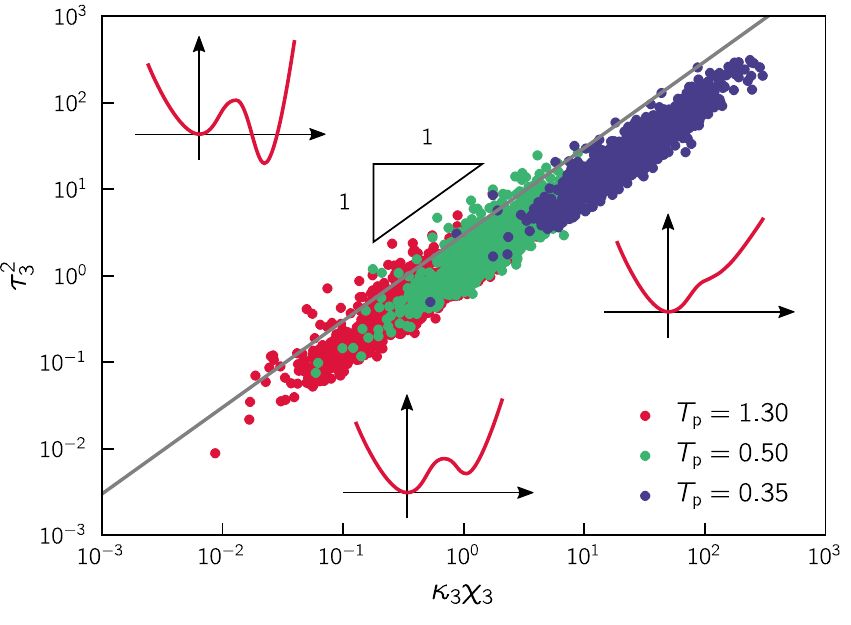}
	\caption{The ensemble of cubic modes follows $\tau_3^2 \sim \kappa_3 \chi_3$, independent of glass stability. The continuous line marks the bound given by \autoref{foo01}, at which the mode --- in the fourth-order expansion of the potential energy --- is a symmetric double-well potential. Modes that fall below this bound are either single-well potentials, or have a second minimum that has a higher energy than the inherent state. Modes that exceed this bound have a second minimum with a lower energy than the inherent state (see illustrations). In our ensemble of solids, only 6.6\% ($T_{\text{p}} = 1.30$), 4.6\% ($T_{\text{p}} = 0.50$), and 0.5\% ($T_{\text{p}} = 0.35$) of cubic modes exceeds this bound. In \autoref{fig:appendix2}\plotlabel{c} (in \Cref{app:data}), we demonstrate that $\chi_3$ is independent of $\kappa_3$, which together with the data in this figure demonstrates that $\tau_3 \sim \sqrt{\kappa_3} \sim \omega_3$, which is central to the predictions that follow in the next section.}
	\label{fig:tau_vs_kappa_chi}
\end{figure}

We therefore scatter plot in \autoref{fig:tau_vs_kappa_chi} the third-order coefficients $\tau_3$ of cubic modes vs.~the product $\kappa_3\chi_3$, calculated for low-lying cubic modes in very stable (blue circles), intermediately stable (green circles) and very unstable (red circles) glasses of $N = 8000$ particles. We find that while the inequality (\ref{foo01}) is not strictly satisfied --- and less so for less stable glasses --- the \emph{entire} set of cubic modes follows
\begin{equation}\label{eq:epic_relation}
\tau_3^2 \sim \kappa_3\chi_3,
\end{equation}
independent of glass stability. This is one of the key results of this work.

Since $\chi_3$ is independent of frequency, as demonstrated in \autoref{fig:appendix2}(c) in \Cref{app:data}, we conclude that cubic modes universally follow 
\begin{equation}\label{foo03}
\tau_3 \sim \omega_3,
\end{equation}
independent of spatial dimension, glass preparation protocol, or microscopic details.

We demonstrate in \autoref{fig:appendix2}(d) of \Cref{app:data} that indeed \autoref{eq:epic_relation} holds in 2D, 3D, and 4D. Finally, we reinforce our statement that only cubic modes accurately capture the third-order coefficient of soft spots in \autoref{fig:appendix2}(e), by demonstrating that the ensemble of quartic modes does \emph{not} follow the relation $\tau_4^2 \sim \kappa_4 \chi_4$.

\subsection{Implications on the statistical properties of the potential energy landscape}

The universal linear scaling between $\tau_3$ and $\omega_3$ (see \autoref{foo03}, \autoref{fig:tau_vs_kappa_chi}, \autoref{fig:appendix2}(c), and discussion in preceding subsection) has two major implications, spelled out here. We focus the discussion exclusively on cubic modes, and therefore omit the subscript `3' in what follows.

\subsubsection{Potential energy barrier distribution}
In the limit $\omega \to 0$, the magnitude $\barrier$ of potential energy barriers that surround a glassy inherent state scales with the coefficients of the NQEs associated with those barriers as \cite{plastic_modes_prerc}
\begin{equation}
\barrier \sim \kappa^3/\tau^2 \sim \omega^4\,,
\end{equation}
where we have used \autoref{foo03}. We further recall that the distribution of cubic modes' frequencies was argued in \autoref{subsec:energetics} to follow the universal ${\cal D}(\omega) \sim \omega^4$ law. Combining these two statements, we obtain a prediction for the distribution of potential energy barriers in the limit $\barrier \to 0$ by a simple transformation of variables, that reads
\begin{equation}\label{foo04}
p(\barrier) \sim \barrier^{1/4}\,.
\end{equation}
The universality of \autoref{foo03} and of the form of ${\cal D}(\omega)$ imply that the distribution of potential energy barriers should universally follow the $\barrier^{1/4}$ law in any model glass, independent of dimension, microscopic details or preparation protocol\footnote{The assumptions leading to \autoref{foo04} have been validated in 2D, 3D, and 4D (see \cite{modes_prl_2018} for the universality of $\mathcal{D}(\omega) \sim \omega^4$, and \autoref{fig:appendix2}(d) for the universality of \autoref{eq:epic_relation}). Preliminary evidence suggests that the universal form of $\mathcal{D}(\omega)$ persists in even higher dimensions \cite{shimada_high_dim}, and we speculate that this might also be the case for \autoref{foo04}. } --- as long as the glasses considered are quenched from a melt.\footnote{A counterexample are the glasses discussed in \cite{fsp} which are not quenched from a melt, and feature a gapped ${\cal D}(\omega)$, and therefore presumably also a gapped $p(\barrier)$} We are currently unaware of numerical data that confirms or refutes our prediction \autoref{foo04}. We note that methods to find energy barriers between inherent states do exist \cite{jonsson1998nudged, henkelman_neb, barkema1998art, cances2009some, mousseau2012art}, but they are computationally expensive and not exhaustive.

\subsubsection{Strain intervals before the first plastic instability}
Much attention has been devoted to understanding the statistics of strain intervals between subsequent plastic instabilities that occur during athermal, quasistatic deformation of computer glasses \cite{lemaitre2004_avalanches,lemaitre2006_avalanches,yielding_rapid_pre_2010,salerno_robbins,jie1,sri_and_itamar_pre_2015,MW_prx_mean_field_yielding,sri_nat_comm_2017,Corrado_strain_intervals_pre_2018,barrat_elastic_branches_avalanches_2019}. In particular, the distribution $p(x) \sim  x^\theta$ of local relative strain thresholds $x$, its finite-size manifestations, and its evolution with shear strain, have been recently debated \cite{sri_and_itamar_pre_2015,MW_prx_mean_field_yielding}. 

Here we recall a result put forward in \cite{micromechanics2016}, that states that the additional local shear deformation $x$ necessary to destabilize a strain-coupled cubic mode follows
\begin{equation}\label{foo06}
x \sim \frac{\omega^4}{\tau \nu},
\end{equation}
where $\nu$ is the same shear-strain-coupling parameter associated with the NQE $\unitmode{\pi}$ as defined after \autoref{foo05}. In \autoref{fig:appendix2}(b) of \aref{app:data} we show that the statistics of the shear-strain-coupling parameter are independent of frequency; therefore, based on \autoref{foo03} and \autoref{foo06} we expect $x \sim \omega^3$. Together with \autoref{foo03} and the universal distribution ${\cal D}(\omega) \sim \omega^4$ of cubic modes, we predict that for as-quenched, isotropic configurations, the distribution of local relative deformation thresholds should follow
\begin{equation}\label{eq:p_of_x}
p(x) \sim x^{2/3}\,,
\end{equation}
i.e.~we predict that $\theta = 2/3$, independent of spatial dimension, microscopic details, or glass preparation protocol. We note that the mean-field theory put forward in \cite{MW_prx_mean_field_yielding} predicts $\theta = 1/2$ for as-quenched, isotropic configurations, while elasto-plastic models feature $\theta \approx 0.6$ in 2D, and $\theta \approx 0.4$ in 3D \cite{jie1}.

The strain interval up to a first plastic instability in a system of size $N$ is now straightforwardly predicted by assuming that it is controlled by the minimal member $x_{\text{min}}$ out of a population of size $\propto  N$ of strain thresholds $x$, namely (see also \cref{foot:weibull})
\begin{equation}\label{eq:xmin}
x_{\text{min}} \sim N^{-\frac{1}{1+\theta}}\sim N^{-\frac{3}{5}}\,,
\end{equation}
in good agreement with \cite{yielding_rapid_pre_2010} where a dimension-independent exponent of -0.62 was measured for the $N$ dependence of $x_{\text{min}}$, and in excellent agreement with \cite{Corrado_strain_intervals_pre_2018} where an exponent of approximately $-0.6$ was found for 2D computer glasses. Other numerical results in 3D feature protocol-dependent exponents ranging between 0.3 and 0.6 \cite{MW_theta_and_omega}. We note that $x_{\text{min}}(N)$ is expected to exhibit strong finite-size effects in stable glasses, because small samples do not feature enough soft quasilocalized excitations for the extreme-value scaling argument \autoref{eq:xmin} to hold, as discussed at length in \cite{Corrado_strain_intervals_pre_2018}. In addition to this, there is another independent finite-size effect that can affect the exponent $\theta$, and therefore the scaling of $x_{\text{min}}(N)$. In \cite{lerner2019finite}, it was shown that when the system size is too small (smaller than the typical core size of QLMs), the density of states of quasilocalized excitations follows
\begin{equation}
	\mathcal{D}(\omega) \sim \omega^{\beta},
\end{equation}
with $\beta$ slightly smaller than 4. This effect is especially noticeable in poorly quenched glasses, which possess quasilocalized excitations with relatively larger disordered cores. In this case, we predict
\begin{equation}
	\theta = (\beta - 2)/3
\end{equation}
(cf.~\autoref{eq:p_of_x}), and
\begin{equation}
	x_{\text{min}} \sim N^{-3 / (1 + \beta)},
\end{equation}
i.e.~we predict a decay slightly stronger than \autoref{eq:xmin} for smaller systems.

\section{Discussion and outlook}\label{sec:discussion}

In this work, we have shown that NQEs are true representatives of soft spots in glasses, unlike the widely used harmonic modes, that --- due to their orthogonality constraint --- always feature a degree of hybridization with other low-energy excitations. We have further shown that a particular class of NQE, termed cubic NQE, reveals universal properties of the potential energy landscape, from which we derive a prediction for the distribution of the lowest energy barriers separating inherent states, and a prediction for the distribution of local strain thresholds to plastic instability.

The practical utility of the nonlinear formulation of soft spots depends crucially on whether all (or most) of the low-energy solutions to the defining equations (\autoref{eq:quartic_def} for quartic and \autoref{cubic_modes_equation} for cubic modes) can be found. This is a computational challenge that we take up in part II of this series of papers, \geert{where we present a method} to find the vast majority of low-lying NQEs --- and, hence, an approximation to the full distribution of soft spots. This opens up several avenues of research to further our understanding of fundamental problems in glass physics, such as the anomalous thermodynamic and transport properties of glasses, yielding, and dynamics in the viscous liquid regime (see citations in the introduction), some of which will be extensively discussed in the next papers in this series.

Another important future research direction is the study of the properties and statistics of NQEs near the \emph{unjamming} transition \cite{ohern2003, liu_review, van_hecke_review, liu2011jamming}. This transition marks the loss of rigidity of the system upon decreasing the degree of connectedness (measured by the excess coordination $\delta z$) of the underlying network of strong interactions below a critical value ($\delta z = 0$), which can, for example, be achieved by decompressing a packing of repulsive soft spheres.
Near this transition, the potential energy landscape becomes highly rugged and hierarchical, and the displacement directions that connect nearby energy minima become delocalized \cite{parisi_fractal, scalliet2019nature}; in contrast, the systems studied in this paper are far from unjamming, and feature simpler energy landscapes, with localized `hopping' between energy minima \cite{scalliet2019nature}. We expect that NQEs will still properly represent soft harmonic QLMs near the unjamming transition. This would imply that the size of their disordered core will diverge as $1 / \sqrt{\delta z}$ \cite{atsushi_core_size_pre}, and their characteristic frequency scale --- which is expected to follow the bulk average of the frequency of the system's response to a local force dipole \cite{cge_paper, pinching_2019} --- will scale as $\delta z$ \cite{new_variational_argument_epl_2016}.

It is important to stress that our prediction for the distribution of local strain thresholds to a plastic instability $p(x) \sim x^{\theta}$, with $\theta = 2/3$, only holds in the \emph{isotropic state}, i.e.~shear strain $\gamma=0$, where the relations $\tau_3 \sim \omega_3$ and $\mathcal{D}(\omega) \sim \omega^4$ hold. At finite $\gamma$ the relation $\tau_3 \sim \omega_3$ must break down, because destabilizing cubic modes' third-order coefficients $\tau_3$ were shown to remain constant upon approaching the instability strain \cite{micromechanics2016}, whereas their frequency vanishes. Moreover, it was recently shown that under shear, $\mathcal{D}(\omega) \sim \omega^{4}$ does not always hold, but instead an exponent smaller than 4 is observed for well-annealed glasses \cite{MW_theta_and_omega}.

This is consistent with recent numerical investigations which have highlighted a non-monotonic behavior of $\theta$ as a function of $\gamma$. In particular, it was shown that $\theta$ first decreases as a function of $\gamma$, before increasing again upon approaching the yielding transition, and finally reaching a plateau value $\theta\approx 1/2$ in the steady state regime \cite{ yielding_rapid_pre_2010, sri_and_itamar_pre_2015, MW_theta_and_omega}. Furthermore, it was reported that the decrease of $\theta$ at intermediate strain is protocol-dependent and is significantly stronger for better-annealed glasses \cite{ sri_and_itamar_pre_2015, barrat_elastic_branches_avalanches_2019}. 
Directly measuring the evolution of $\mathcal{D}(\omega)$ and the joint distribution $P(\omega_3,\tau_3)$ under shear, for various degrees of annealing, is therefore a highly promising direction of future research.


\acknowledgments

We warmly thank Corrado Rainone and Eran Bouchbinder for fruitful discussions. E.~L.~acknowledges support from the Netherlands Organisation for Scientific Research (Vidi grant no.~680-47-554/3259). 
We are grateful for the support of the Simons Foundation for the ``Cracking the Glass Problem Collaboration" Awards No.~348126 to Sid Nagel (D.~Richard).

\appendix
\section{Tensorial notation}\label{app:tensor}
To aid readability, in this work we omit particle indices (denoted here by roman letters, e.g.~$i = 1, \dots, N$) and Cartesian (spatial) indices (denoted here by greek letters, e.g.~$\alpha = x, y, z$) from all tensorial and vectorial quantities. For example, the $N\dbar$-dimensional vector $\mode{x}$ should be understood as $x_i^{\alpha}$, and the tensor $\mode{U}^{(4)} \equiv \frac{\partial U}{\partial \mode{x} \partial \mode{x} \partial \mode{x} \partial \mode{x}}$ should be understood as $\frac{\partial U}{\partial x_i^{\alpha} \partial x_j^{\beta} \partial x_{k}^{\gamma} \partial x_{l}^{\delta}}$. We denote single, double, triple and quadruple contractions by $\cdot$, $\doubleCdot$, $\tripleCdot$ and $\quadCdot$, respectively. For example, the expression $\mode{U}^{(4)} \tripleCdot \mode{z} \mode{z} \mode{z}$ should be understood as
	$\frac{\partial U}{\partial x_i^{\alpha} \partial x_j^{\beta} \partial x_{k}^{\gamma} \partial x_{l}^{\delta}} z_i^{\alpha} z_j^{\beta} z_k^{\gamma}$, where the summation convention applies.

\section{NQE analysis}\label{app:nqe}
For all models used, we obtain a collection of low-energy harmonic QLM and corresponding quartic and cubic modes as follows. For each system, we first find the lowest eigenvalue and eigenmode of the Hessian matrix (not including the $\dbar$ translational modes) using ARPACK \cite{arpack}. Then, we use the lowest eigenmode as an initial vector to locally minimize the quartic energy function (\autoref{eq:quartic_energy}) to obtain the corresponding quartic NQE. Finally, we use the quartic mode as an initial vector to locally minimize the cubic energy function (\autoref{eq:cubic_energy}) to obtain the corresponding cubic mode. This procedure results in one harmonic, one quartic and one cubic mode for each system. 

For the 3D POLY model, we apply an additional procedure to find a low-energy cubic NQE in each system. Besides the quartic NQE, we consider as initial vectors the six \emph{non-affine velocities} \cite{athermal_elasticity} that result from imposing a simple and pure shear deformation in the XY-, XZ-, and YZ-plane. From these seven initial guesses, we select the cubic NQE with the lowest stiffness. 

To minimize the cubic and quartic energy functions, we use the macopt conjugate-gradient minimizer \cite{macopt_cg}, modified such that the variable vector $\mode{z}$ is always kept normalized.

The two quartic modes shown in \autoref{fig:hybridization}\plotlabel{b} were found as follows. The first quartic mode, shown in black, was found by using the harmonic double-core QLM $\unitmode{\psi}$ (see panel \plotlabel{a}) as an initial condition to minimize the quartic energy function. The second quartic mode, shown in green, was found by using the difference between $\unitmode{\psi}$ and the first quartic mode as an initial condition.

\section{Scaling relation between participation and fourth-order expansion coefficient}\label{app:chi_vs_Ne}
In this Appendix we explain the scaling relation $\chi_4 \sim (Ne_4)^{-1}$, that holds independently of system size and glass preparation protocol, as shown in \autoref{fig:chi_vs_e}. Here, $\chi_4 \equiv \chi(\unitmode{\pi}_4)$ is the fourth-order expansion coefficient (\autoref{eq:chi_def}), and $e_4 \equiv  e(\unitmode{\pi}_4)$ is the participation ratio (\autoref{eq:e_def}), associated with quartic modes $\unitmode{\pi}_4$.

First, observe that for any normalized mode, we have
\begin{equation}\label{eq:eN_inv}
	\frac{1}{N e(\unitmode{z})} = \sum_i^N \left( \unitmode{z}_i \cdot \unitmode{z}_i \right)^2.
\end{equation}
We now argue that $\chi(\unitmode{z}) \equiv \mode{U}^{(4)} \quadCdot \unitmode{z} \unitmode{z} \unitmode{z} \unitmode{z}$ has \geert{a similar structure}.
For example, for radially-symmetric pairwise potentials, we have
\begin{eqnarray}
\chi(\unitmode{z}) 
&  = & \sum\nolimits_{\expval{ij}} \left( \sFrac{\varphi''''_{ij}}{r_{ij}^4}  -  \sFrac{6\varphi'''_{ij}}{r_{ij}^5}  +  \sFrac{15\varphi''_{ij}}{r_{ij}^6}  -  \sFrac{15\varphi'_{ij}}{r_{ij}^7}\right)(\mode{x}_{ij} \cdot \unitmode{z}_{ij})^4 \nonumber \\
& + & 6\sum\nolimits_{\expval{ij}} \left( \sFrac{\varphi'''_{ij}}{r_{ij}^3} -  \sFrac{3\varphi''_{ij}}{r_{ij}^4}  +  \sFrac{3\varphi'_{ij}}{r_{ij}^5}\right) (\mode{x}_{ij} \cdot \unitmode{z}_{ij})^2(\unitmode{z}_{ij} \cdot \unitmode{z}_{ij})  \nonumber \\
& + &  3\sum\nolimits_{\expval{ij}}  \left( \sFrac{\varphi''_{ij}}{r_{ij}^2} - \sFrac{\varphi'_{ij}}{r_{ij}^3} \right)  (\unitmode{z}_{ij}\cdot\unitmode{z}_{ij})^2,
\end{eqnarray}
\geert{where the sums run over $O(N)$ interacting pairs $\langle i j \rangle$. Here, $\mode{x}_{ij} \equiv \mode{x}_j - \mode{x}_i$ and $\unitmode{z}_{ij} \equiv \unitmode{z}_j - \unitmode{z}_i$. Aside from prefactors and contractions with $\mode{x}_{ij}$, every term scales as $|\unitmode{z}_{ij}|^4$. So $\chi(\unitmode{z}) \sim \sum_{\expval{ij}} |\unitmode{z}_{ij}|^4$ functions rather like $1 / Ne(\unitmode{z})$, except that $\unitmode{z}_i$ is replaced by $\unitmode{z}_{ij}$ (cf.~\autoref{eq:eN_inv}), which explains the observed scaling $\chi_4 \sim (Ne_4)^{-1}$. Furthermore, it explains why quartic modes, in addition to having a low participation, so effectively disentangle localized soft spots from a phonon background: terms that scale as $|\unitmode{z}_{ij}|^4$ are extremely small for phonon-like fields, meaning that those fields are suppressed in the minima of the associated nonlinear cost function (cf.~\autoref{eq:quartic_energy}).\footnote{The same argument holds for cubic modes, whose cost function features the third-order coefficient $\tau$, which contains terms that scale as $|\mode{z}_{ij}|^3$.} This suggests that minima of a cost function in which $\chi(\unitmode{z})$ is replaced by the purely geometric factor $\sum_{\expval{ij}} (\unitmode{z}_{ij} \cdot \unitmode{z}_{ij})^2$, i.e.
\begin{equation}
	C(\mode{z}) = \frac{\left( \dyn \doubleCdot \mode{z}\mode{z} \right)^2}{\sum_{\expval{ij}} (\mode{z}_{ij} \cdot \mode{z}_{ij})^2},
\end{equation}
will be similar to quartic modes.}

\section{Derivation of the expression for $\mode{\Delta}$}\label{app:scaling}
In this Appendix, we present a derivation of \autoref{eq:delta} for the difference vector of the quartic NQE (of stiffness $\omega_4^2$) and the harmonic QLM (of stiffness $\omega_2^2$), denoted by $\mode{\Delta} \equiv \unitmode{\pi} - \unitmode{\psi}$. This derivation is very similar to the one presented in \cite{micromechanics2016} for the difference between the cubic NQE and the destabilizing harmonic mode near a plastic instability under an imposed shear deformation.

Recall that the stiffness associated with a particular displacement field $\mode{z}$ is defined as
\begin{equation}
	\kappa(\mode{z}) = \frac{\dyn \doubleCdot \mode{z}\mode{z}}{\mode{z}\cdot\mode{z}}.
\end{equation}
For the analysis that follows, we require the first and second derivatives
\begin{align}
	\pdv{\kappa}{\mode{z}} &= \frac{2}{\mode{z}\cdot\mode{z}}\left( \dyn \cdot \mode{z} - \kappa(\mode{z}) \mode{z} \right), \\
	\pdv[2]{\kappa}{\mode{z}} &= \frac{2}{(\mode{z}\cdot\mode{z})^2}\bigg[ (\mode{z}\cdot\mode{z})\dyn - 4 (\dyn \cdot \mode{z}) \otimes \mode{z} \\
														&- (\dyn \doubleCdot \mode{z}\mode{z}) \identity + 4\kappa(\mode{z}) \mode{z}\otimes\mode{z}  \bigg].\nonumber
\end{align}
Here, $\identity$ denotes the identity matrix. We will now leverage the fact that the harmonic QLM represents the system's displacement direction of smallest stiffness, so that
\begin{equation}
	\eval{\pdv{\kappa}{\mode{z}}}_{\unitmode{\psi}} = 0.
\end{equation}
We write down the first-order Taylor expansion of the function $\pdv{\kappa}{\mode{z}}$ around $\mode{z} = \unitmode{\psi}$ to approximate
\begin{equation}
	\eval{ \pdv{\kappa}{\mode{z}} }_{\unitmode{\pi}} \simeq \eval{\pdv[2]{\kappa}{\mode{z}}}_{\unitmode{\psi}} \cdot \mode{\Delta}.	
\end{equation}
Inverting in favor of $\mode{\Delta}$, and using \autoref{eq:quartic_def} to simplify yields
\begin{equation}
	\mode{\Delta}  \simeq \omega_4^2 \left( \dyn - \omega_2^2 \identity \right)^{-1} \left(\frac{\mode{U}^{(4)}\tripleCdot \unitmode{\pi}_4 \unitmode{\pi}_4 \unitmode{\pi}_4}{\chi_4} - \unitmode{\pi}_4 \right),
\end{equation}
which, once $\left( \dyn - \omega_2^2 \identity \right)^{-1}$ is written in diagonalized form, is equal to \autoref{eq:delta}.

\section{Supporting data}\label{app:data}

\geert{In this Appendix, we have collected data that supports various claims made in the main text.}

\begin{figure}[htpb]
	\centering
	\includegraphics[]{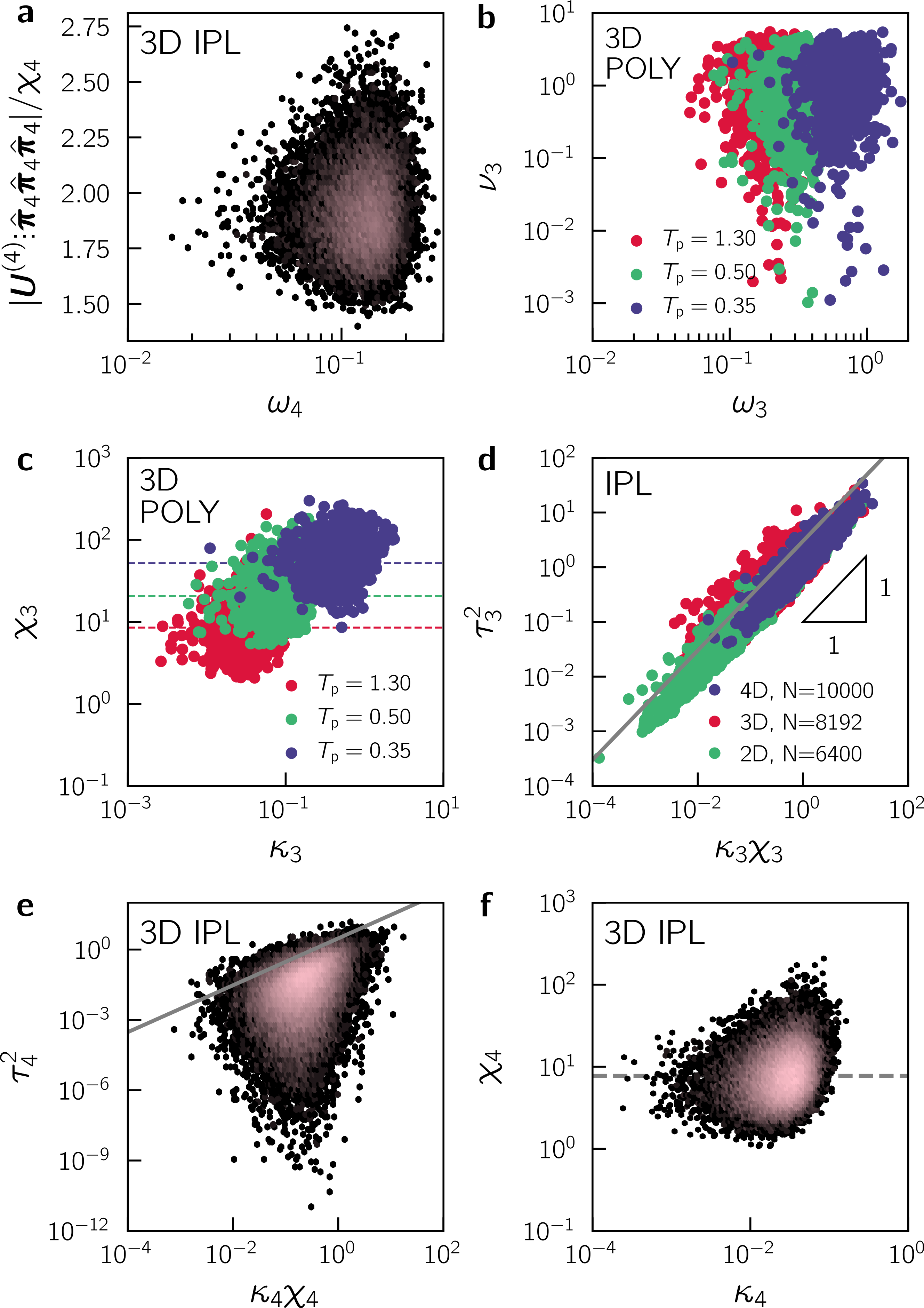}
	\caption{(a) The magnitude of the field $\mode{U}^{(4)}\tripleCdot \unitmode{\pi}_4 \unitmode{\pi}_4 \unitmode{\pi}_4 / \chi_4$ is of order $O(1)$, independent of $\omega_4$. (b) The shear-strain-coupling parameter $\nu_3 \equiv \frac{\partial^2 U}{\partial\gamma\partial\mode{x}} \cdot \unitmode{\pi}_3$ associated with cubic modes $\unitmode{\pi}_3$ is independent of $\omega_3$. (c) The fourth-order coefficient of cubic modes $\chi_3$ is independent of $\omega_3$. (d) The entire ensemble of cubic modes follows the relation $\tau_3^2 \sim \kappa_3 \chi_3$ in 2D, 3D, and 4D. (e) The ensemble of quartic modes does \emph{not} follow this relation. (f) The fourth-order coefficient of quartic modes $\chi_4$ is independent of $\omega_4$. In panels (a, e, f), the data is presented as a 2D histogram, where each bin is colored according to the number of hits in that bin (the color-coding is the same as in \autoref{fig:why_cubic_modes}).}
	\label{fig:appendix2}
\end{figure}

\FloatBarrier

\bibliography{references_lerner}

\end{document}